\documentclass[sigconf,nonacm]{acmart}

\usepackage{xspace}
\newcommand{\sys}{{\textit{AFLGopher}}\xspace}
\newcommand{\PP}[1]{
\noindent{\bf \IfEndWith{#1}{.}{#1}{#1.}}
}

\usepackage{threeparttable}
\usepackage{graphicx}
\usepackage{listings}
\usepackage{subcaption}
\usepackage{verbatim}
\usepackage{longtable}
\usepackage{makecell}
\usepackage{float}
\usepackage{tcolorbox}

\definecolor{codegreen}{rgb}{0,0.6,0}
\definecolor{codegray}{rgb}{0.5,0.5,0.5}
\definecolor{codepurple}{rgb}{0.58,0,0.82}
\definecolor{backcolour}{rgb}{0.95,0.95,0.92}
\definecolor{emphcolor}{rgb}{0.58,0,0.29} % Color for emphasized items
\definecolor{highlight}{rgb}{0,0,1}
\definecolor{highlight2}{rgb}{1,0.64,0}

\definecolor{packagecolor}{rgb}{0.5, 0.0, 0.5}  % Purple
\definecolor{descriptioncolor}{rgb}{0.0, 0.5, 0.5} % Teal
\definecolor{bannedcolor}{rgb}{0.85, 0.1, 0.1} % Red

\usepackage{tcolorbox}

\newtcolorbox{TABox}{colback=white, colframe=black, boxrule=0.8pt, sharp corners}

\usepackage{diagbox}
\usepackage{tikz}
\newcommand*{\circled}[1]{\lower.7ex\hbox{\tikz\draw (0pt, 0pt)%
    circle (.5em) node {\makebox[1em][c]{\small #1}};}}

\usepackage{markdown}
\usepackage{bbding}
\usepackage{wrapfig}
\usepackage{soul}

\usepackage{tabularx}
\usepackage{array}

\usepackage{multirow}

\markdownSetup{fencedCode = true}

\usepackage{xcolor}
\usepackage{color}

\AtBeginDocument{%
  }

\begin{document}

\title{AFLGopher: Accelerating Directed Fuzzing via Feasibility-Aware Guidance}

\author{Weiheng Bai}
\email{bai00093@umn.edu}
\affiliation{%
  \institution{University of Minnesota - Twin Cities}
  \city{Minneapolis}
  \state{MN}
  \country{USA}
}

\author{Kefu Wu}
\email{wu000380@umn.edu}
\affiliation{%
  \institution{University of Minnesota - Twin Cities}
  \city{Minneapolis}
  \state{MN}
  \country{USA}
}

\author{Qiushi Wu}
\email{qiushi.wu@ibm.com}
\affiliation{%
  \institution{IBM T.J. Watson Research Center}
  \city{Yorktown Height}
  \state{NY}
  \country{USA}
}

\author{Kangjie Lu}
\email{kjlu@umn.edu}
\affiliation{%
  \institution{University of Minnesota - Twin Cities}
  \city{Minneapolis}
  \state{MN}
  \country{USA}
}

\renewcommand{\shortauthors}{Trovato et al.}


\begin{abstract}
Directed fuzzing is a useful testing technique that aims to
efficiently reach target code sites in a program.
It has a wide range of applications, such as assessing the severity of vulnerabilities, confirming bugs found by static analysis, reproducing
existing bugs, and testing code changes.
The core of directed fuzzing is the guiding mechanism that 
directs the fuzzing to the specified target.
A general guiding mechanism adopted in existing directed
fuzzers is to calculate
the control-flow distance between the current progress and the target,
and use that as feedback to guide the directed fuzzing.
A fundamental problem with the existing guiding mechanism
is that the distance calculation is \emph{feasibility-unaware}. 
For instance, it always assumes that the two branches
of an \texttt{if} statement have equal feasibility, which
is likely not true in real-world programs and would inevitably
incur significant biases that mislead directed fuzzing.

In this work, we propose \emph{feasibility-aware} directed 
fuzzing named \sys.
Our new feasibility-aware distance calculation provides
pragmatic feedback to guide directed fuzzing to reach
targets efficiently. We propose new techniques to address
the challenges of feasibility prediction. Our new classification method
allows us to predict the feasibility of all branches based
on limited traces, and our runtime feasibility-updating
mechanism gradually and efficiently improves the prediction
precision.
We implemented \sys and compared \sys with state-of-the-art directed fuzzers including AFLGo, enhanced AFLGo, WindRanger, BEACON and SelectFuzz.
\sys is 3.76$\times$, 2.57$\times$, 3.30$\times$, 2.52$\times$ and 2.86$\times$ faster than AFLGo, BEACON, WindRanger, SelectFuzz and enhanced AFLGo, respectively, in reaching targets.
\sys is 5.60$\times$, 5.20$\times$, 4.98$\times$, 4.52$\times$, and 5.07$\times$ faster than AFLGo, BEACON, WindRanger, SelectFuzz and enhanced AFLGo, respectively, in triggering known vulnerabilities.
\end{abstract}

%   <concept_desc>Security and Privacy, ML and AI applications to security and privacy</concept_desc>

% \ccsdesc[500]{Security and Privacy~ML and AI applications to security and privacy}

% %%
% %% Keywords. The author(s) should pick words that accurately describe
% %% the work being presented. Separate the keywords with commas.
% \keywords{Fuzzing, Directed Fuzzing, Machine Learning, Deep Learning}

\maketitle

\section{Introduction}

Directed fuzzing aims to generate inputs that steer program execution toward specific target locations — such as newly patched code, suspicious functions, or frames in crash stack traces. It has become a foundational technique in modern software security testing, widely used for patch validation~\cite{b19, b21}, vulnerability reproduction~\cite{b7}, and exploitability assessment~\cite{b26}.

Unlike coverage-based fuzzers that explore programs indiscriminately, directed fuzzers focus on specific destinations. Their goal is two-fold: (1) \textit{reach the target location as quickly and reliably as possible}, and (2) \textit{once reached, fuzz around the target to expose potential vulnerabilities}. This \textbf{reach-then-trigger} model enables effective use of limited fuzzing resources and is critical in scenarios like crash reproduction and patch testing.

To guide the search, most directed fuzzers build on coverage-guided fuzzers like AFL~\cite{b1}, incorporating distance metrics to prioritize seeds closer to the target. AFLGo~\cite{b14}, the seminal work in this space, formulates the problem as a global optimization: it computes control-flow graph (CFG) distances from each basic block to the target and uses simulated annealing to allocate more energy to closer seeds. Hawkeye~\cite{b2} and WindRanger~\cite{b54} improve this approach by incorporating indirect-call resolution and data-flow sensitivity, respectively. However, all these systems share a critical limitation: they are \textit{feasibility-unaware}. They treat CFG distance as a proxy for reachability and assume all branches are equally likely to be taken.

\PP{Why prior work falls short}
In practice, control-flow branches vary greatly in their likelihood of being executed. Consider the example in Listing~\ref{lst:auth}, where a vulnerability lies inside a function behind a rarely satisfied token check. A fuzzer like AFLGo may repeatedly attempt to reach this target because it is “close” in terms of graph distance, despite the branch being highly improbable. This misalignment causes wasted effort and poor performance.

\begin{figure}[htbp]
\centering
\begin{lstlisting}
void processRequest(char* token) {
    if (strcmp(token, "admin-token") != 0) {
        return; 
        // Reject request if token is not an exact match
    }
    dangerousFunction(); 
    // Target Position: Function that needs to be tested
    
}\end{lstlisting}
    \caption{A rare condition guards access to a vulnerability.}
    \label{lst:auth}
\end{figure}

Recent learning-based fuzzers attempt to address this differently. DeepGo~\cite{b70} and HyperGo~\cite{b67} learn mutation and path-transition probabilities to better guide input generation, but they do not model the feasibility of specific branches or generalize across structurally similar conditions. BEACON~\cite{b61} prunes statically unreachable paths using symbolic reasoning, but lacks probabilistic prioritization and cannot adapt at runtime. These systems either treat feasibility as binary or focus only on mutation-level feedback.

\PP{Feasibility-Aware Directed Fuzzing}
We propose \textbf{\sys}, the first directed fuzzer that augments traditional distance-based guidance with explicit \emph{branch feasibility estimation}. 
Rather than assuming that all control-flow paths are equally likely to be taken, \sys learns the empirical reachability of branches through runtime execution traces, semantic clustering, and sequence modeling. 
These feasibility scores are then embedded directly into the distance metric used to prioritize fuzzing seeds—shifting guidance from structural proximity to a more principled objective: \textbf{expected success in reaching the target}.
This reformulates directed fuzzing from a static optimization problem over control-flow structure into a \emph{probabilistic optimization problem} over executable paths, aligning fuzzing effort with real-world execution likelihood.

To realize feasibility-aware guidance, \sys must determine the likelihood that a given control-flow edge is taken during fuzzing. 
In practice, not all program statements require this modeling. 
If a statement has only one successor (e.g., straight-line code), its outgoing edge has a feasibility of 100\%. Hence, we focus exclusively on statements with multiple possible successors—referred to in this paper as \emph{branch statements}.

\PP{Definition of branch statement} 
A branch statement is any program point where execution may proceed along more than one distinct path, depending on input or runtime state. 
In our work, we classify three types of branch statements: (1) \textbf{conditional statements} such as \texttt{if} and \texttt{switch}, (2) \textbf{indirect calls or jumps}, and (3) \textbf{return statements}. However, since the destination of a return is statically determined by the call site, its feasibility is trivially 100\%. Therefore, \sys focuses on estimating feasibility for the first two classes: conditional branches and indirect calls.

By modeling the likelihood of taking each branch at these control points, \sys can accurately estimate the true cost of reaching a target, prioritize high-probability paths, and avoid wasting effort on structurally short but semantically unreachable ones.

% \PP{Challenges and key techniques}
% Building feasibility-aware directed fuzzing presents three main challenges. First, most branches remain uncovered during fuzzing, making it difficult to infer feasibility from limited data (\textbf{C1}). Second, feasibility changes over time as new inputs are discovered, requiring efficient online adaptation (\textbf{C2}). Third, different branch types (e.g., indirect calls vs. conditionals) have distinct structural and semantic behaviors (\textbf{C3}).

% To address these, \sys introduces: (i) a semantics-aware clustering method to propagate feasibility from covered to uncovered conditional branches; (ii) an LSTM-based model for predicting indirect-call reachability from function-call sequences; and (iii) a lightweight runtime update mechanism that monitors error rates and incrementally refines predictions.

\PP{Challenges} 
To realize \sys, we face several challenges.
To predict the feasibility of the branches, we need ground-truth data---the traces collected through dynamic analysis.
However, a fundamental problem with dynamic analysis
is that it cannot be comprehensive, with a very limited
coverage rate.
Most branches will not be covered and will not have any traces.
It is challenging to predict the feasibility of all branches based on the limited traces (\textbf{C1}).
Second, fuzzing is a dynamic process; to ensure precision, the prediction should keep improving efficiently as more traces become available.
Without an optimization mechanism, the directed fuzzing will likely stall and fail to reach the targets. Optimization should be a runtime mechanism, so ensuring efficiency becomes challenging (\textbf{C2}).

\PP{Our techniques} 
We propose several new techniques to address the challenges mentioned above.
Our first technique is a new classification method that allows us to calculate the feasibility of conditional-statement branches that do not have traces. 
The idea is to predict their feasibility based on the most similar conditional-statement branches that have traces, since semantically similar conditional statements are supposed to share similar feasibility
distribution on their branches.
Our second technique is a new deep-learning-based method that leverages the ability of the Long Short-Term Memory (LSTM) model: perfectly suitable for single-direction sequence-sensitive problems.
This method allows us to take the traces as function-call sequences and predict the feasibility of each candidate's indirect-call target.
Our third technique is a highly efficient runtime-updating mechanism that gradually improves the precision of feasibility prediction. 
The idea is that as the directed fuzzing progresses, we will have more and more traces that can be used to improve the prediction precision.
To ensure efficiency, we further develop on-demand prediction-model updating and incremental deep learning, which triggers updating only when necessary and updates only the relevant parts without retraining the prediction model.

We implement \sys and rigorously evaluate it against several state-of-the-art directed fuzzing tools, including BEACON, WindRanger, SelectFuzz, and AFLGo. 
Recognizing that AFLGo lacks support for indirect call edges, we also develop an enhanced variant, termed AFLGo-ICall, which extends its capabilities to handle indirect calls.

Our experimental results demonstrate that \sys significantly outperforms all baseline fuzzers in reaching target code locations, achieving speedups of 3.76$\times$, 2.57$\times$, 3.30$\times$, 2.52$\times$, and 2.86$\times$ over AFLGo, BEACON, WindRanger, SelectFuzz, and AFLGo-ICall, respectively. Moreover, in reproducing known vulnerabilities, \sys again leads with improvements of 5.60$\times$, 5.20$\times$, 4.98$\times$, 4.52$\times$, and 5.07$\times$ across the same set of baselines.

While direct comparison with DeepGo is not entirely equitable—due to the unavailability of its source code—our detailed evaluation and discussion in~\autoref{sec:eval} substantiate that \sys consistently outperforms DeepGo across shared benchmarks. We attribute this performance gain to \textit{AFLGopher}’s principled guidance mechanism that integrates dynamic feasibility estimation, enabling more efficient navigation of semantically reachable paths in the fuzzing process.

In this paper, we make the following contributions:

$\bullet$ \textbf{Feasibility-aware directed fuzzing.} We introduce the first feasibility-aware distance metric that integrates branch feasibility into directed fuzzing guidance. This enables more accurate feedback and significantly improves efficiency in reaching target code locations.

$\bullet$ \textbf{Techniques for feasibility prediction.} We develop three complementary methods: (1) a semantic classification approach that predicts the feasibility of conditional branches even without direct execution traces, (2) an LSTM-based model that estimates the feasibility of indirect-call targets from limited traces, and (3) a lightweight runtime update mechanism that incrementally refines predictions as fuzzing progresses.

$\bullet$ \textbf{Implementation and evaluation.} We design and implement \sys, a feasibility-aware fuzzer that combines static and dynamic analysis. Our extensive evaluation across real-world benchmarks shows that \sys reduces time-to-target and time-to-exposure by factors of 2.5–5× compared to state-of-the-art directed fuzzers.

$\bullet$ \textbf{Artifact availability.} To foster open science and reproducibility, we will release the full implementation of \sys after acceptance.

\section{Background}

\subsection{Directed Fuzzing: Design and Limitations}

Directed fuzzing guides test generation toward a specified set of target locations in the program — such as recently patched code, a crash site, or a statically flagged vulnerability. Among existing strategies, distance-guided fuzzing has emerged as the dominant design, pioneered by AFLGo~\cite{b14}.

AFLGo models directed fuzzing as an optimization problem: it statically computes the control-flow graph (CFG) distance between basic blocks and the target, and uses this distance to rank seeds. The fuzzer allocates more energy to inputs that appear closer to the target, using simulated annealing to balance exploration and exploitation over time.

Subsequent works, such as Hawkeye~\cite{b2} and WindRanger~\cite{b54}, refine distance modeling by resolving indirect calls or incorporating data-flow dependencies. These systems remain grounded in the assumption that static proximity in the control-flow graph correlates with actual reachability. However, this assumption often breaks down in practice.

Critically, traditional distance-based fuzzers treat all branches and paths as equally feasible — ignoring that many are rarely or never taken under realistic inputs. As a result, fuzzers frequently prioritize syntactically short paths that are semantically unreachable. This gap between control-flow distance and true execution feasibility leads to ineffective exploration and wasted effort, particularly in programs with deep, input-sensitive logic.

\subsection{The Need for Feasibility Awareness}

While control-flow distance provides a convenient structural heuristic for prioritizing fuzzing inputs, it fails to capture the actual difficulty of reaching target code in practice. In many real-world programs, control-flow branches exhibit highly imbalanced feasibility: some are taken almost always, while others are rarely or never executed unless a specific input satisfies stringent conditions.

This discrepancy causes a critical failure mode in existing directed fuzzers. Because they treat all branches as equally viable, they often prioritize seeds that appear structurally close to the target but are routed through infeasible paths — resulting in wasted effort. Our empirical analysis reveals the severity of this problem: in our benchmarks we used in \autoref{sec:eval}, 96\% of branch statements exhibit unequal feasibility across their outgoing edges. Nonetheless, existing directed fuzzers assign equal cost to both sides of a branch, ignoring this disparity.

The impact is substantial. According to the FuzzGuard study~\cite{b30}, over 91.7\% of inputs generated by state-of-the-art directed fuzzers fail to reach the target location. To investigate this further, we manually analyzed execution traces of unreachable inputs and found that many were blocked by low-feasibility branches — such as \texttt{if} statements involving hard-coded tokens, size limits, or allocation failures.

These findings highlight a fundamental insight: \emph{static proximity does not imply reachability}. A path with minimal CFG distance may, in reality, be effectively unreachable within the fuzzing time budget. To address this, we argue that directed fuzzers must shift from structural heuristics to \textbf{feasibility-aware guidance} — incorporating the likelihood that a path can actually be taken under mutation.

This motivates our design of \sys, which integrates empirical feasibility into the core distance metric. In \autoref{sec:motivation}, we present two key observations that ground this design: (1) path priority changes significantly when feasibility is considered, and (2) branch feasibility can be generalized across semantically similar structures.

\subsection{Empirical Insights: Feasibility Influences Path Selection}
\label{sec:motivation}

To evaluate how branch feasibility affects fuzzing guidance, we conduct a detailed analysis of real-world programs and vulnerabilities. Our findings support two key insights: (1) incorporating feasibility changes which paths are prioritized, and (2) semantically similar branches tend to share feasibility characteristics, enabling generalization.

\begin{figure}[ht]
% \centerline{\includegraphics[width=0.55\textwidth]{graphs/trace1.pdf}}
\centerline{\includegraphics[width=0.3\textwidth]{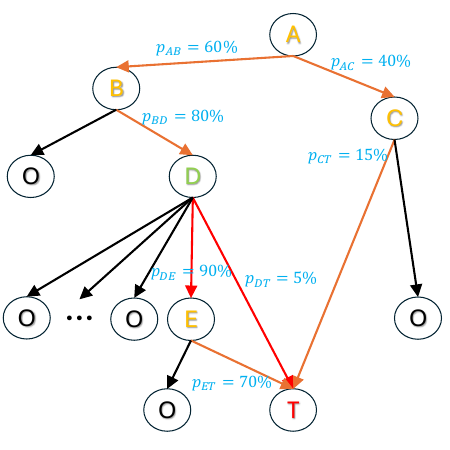}}
\caption{Call traces for exploiting the CVE-2020-11895. The node $T$ represents the target function. The node $A-E$ represents the functions of the traces to the target. Node $O$ represents all functions that are not included in the traces to the target. Node $A, B, C, E$ has a \textsc{if} statement that has two outgoing edges; node $D$ indicates indirect calls and has several potential indirect call sites.}
\label{fig:trace}
\end{figure}

\PP{Feasibility alters prioritization.}
We analyze CVE-2020-11895 in \texttt{libming}, a real-world vulnerability that requires a specific path to reach the buggy function. \autoref{fig:trace} shows three potential call traces leading to the target node $T$, with intermediate functions labeled $A$ to $E$. Several functions contain conditional branches ($A$, $B$, $C$, $E$) or indirect calls ($D$), resulting in different levels of execution feasibility.

Using AFLGo’s static CFG-based distance calculation, Trace 1 (\texttt{A} → \texttt{C} → \texttt{T}) appears shortest and is therefore prioritized. However, our feasibility-aware analysis reveals that Trace 3 (\texttt{A} → \texttt{B} → \texttt{D} → \texttt{E} → \texttt{T})—though longer structurally—has much higher likelihood of successful execution due to more favorable branch conditions and indirect-call targets. By computing the empirical feasibility of each control-flow edge using AFL trace coverage, we quantify this as:

\begin{equation}
    P_{ij} = \frac{\text{Hit}_j}{\text{Hit}_i}
    \label{eq:feasiblity}
\end{equation}

Where $\text{Hit}_i$ is the number of times node $i$ is reached, and $\text{Hit}_j$ is the number of times control flows from $i$ to $j$. Using this metric, \sys reprioritizes traces based on both path structure and actual feasibility, avoiding misleading shortcuts and focusing effort where success is more likely.

\PP{Semantically similar branches exhibit similar feasibility.}
Prior work has demonstrated that semantically similar code fragments tend to exhibit similar behavior or functionality. 
For instance, Pei et al.\cite{pei2022semantic} and Ming et al.\cite{ming2017obfuscation} show that structurally and semantically similar functions yield comparable execution traces, while Tufano et al.~\cite{tufano2019repair} leverage semantic embeddings to transfer program repair patterns across code segments. 
These findings suggest that semantic similarity can serve as a proxy for functional similarity, even across lexically diverse code.

Motivated by this insight, we hypothesize that \emph{semantically similar conditional branches tend to share similar execution feasibility} — that is, the likelihood of triggering a particular branch direction during fuzzing. 
For example, branches guarding resource allocation failures (e.g., \texttt{malloc} or \texttt{xmlAlloc}) or dictionary initialization typically exhibit low feasibility on their failure side, while accepting branches are taken more frequently. 
We posit that this asymmetry is not unique to specific branches but consistent across semantically similar ones throughout the program.

However, as shown in Listing~\ref{lst:similar}, we find that surface-level semantics of the condition itself (e.g., \texttt{if(!inbuffer)} or \texttt{if(out == NULL)}) are insufficient to accurately determine similarity.
Although these two branches perform equivalent checks, the variables involved—\texttt{inbuffer} and \texttt{out}—do not immediately reveal their semantics. To address this, \sys introduces \emph{context-sensitive backward analysis} (\autoref{design:semantics}) to trace each condition variable to its source, capturing more meaningful semantic context, such as whether it originates from a memory allocator.

\begin{figure}[htbp]
\centering
\begin{lstlisting}
void foo1(){
        ...
        inbuffer = malloc(insize);
        ...  
        if(!inbuffer){...}  /*if-1*/
        ...
        Buffer out = (Buffer)malloc(BUFFER_SIZE);
        ...    
        if(out == NULL){...}  /*if-2*/
    }
    \end{lstlisting}
    \caption{Two semantically similar branches guarded by memory allocation checks.}
    \label{lst:similar}
\end{figure}

To validate this, we conducted an empirical study across seven diverse open-source programs spanning domains such as parsing, compression, memory management, and system utilities. 
From these programs, we curated a stratified benchmark of 1,000 conditional branch statements. 
These findings are intended to provide empirical, directional guidance for the feasibility generalization approach employed in \sys. 
A detailed description of our sampling methodology and validation strategy is provided in~\autoref{sec:metrics}.

To cluster the branches, we applied a semantic taxonomy based on three axes (detailed in \autoref{design:semantics}):

\noindent $\bullet$ \textit{Condition variable role}, such as return-value checks (e.g., \texttt{if (malloc(...) == NULL)}), pointer checks, global state flags, and struct field access;

\noindent $\bullet$ \textit{Comparison structure}, including null checks, integer thresholds, range bounds, and bitmask operations;

\noindent $\bullet$ \textit{Usage context}, inferred from identifier names and API calls, such as memory allocation, file I/O, parsing, error propagation, and resource availability.

This clustering process resulted in 46 distinct semantic clusters. All branch labels were independently annotated and reviewed by three experienced graduate-level C/C++ developers (each with 5+ years of experience), and disagreements were resolved by majority vote to ensure consistency and quality.

Each program was fuzzed using AFL, and dynamic execution traces were collected to measure the feasibility of each branch — defined as the relative frequency of its outgoing edges being taken. To assess whether semantic similarity aligns with feasibility, we compared our manual semantic clusters with feasibility-based groupings using the Adjusted Rand Index (ARI). Our clustering achieved an ARI of 0.91, indicating excellent alignment. In contrast, random clusterings of the same data yielded an average ARI of 0.29.

These results strongly support our hypothesis: \textbf{branches with similar semantics tend to share similar feasibility}. This empirical observation validates the use of semantic clustering in \sys to generalize feasibility estimates across semantically similar but previously unexplored branches, enabling more informed and efficient fuzzing guidance.

\section{Overview of \sys}
\label{sec:overview}

\begin{figure*}[!ht]
\centerline{\includegraphics[width=1\textwidth]{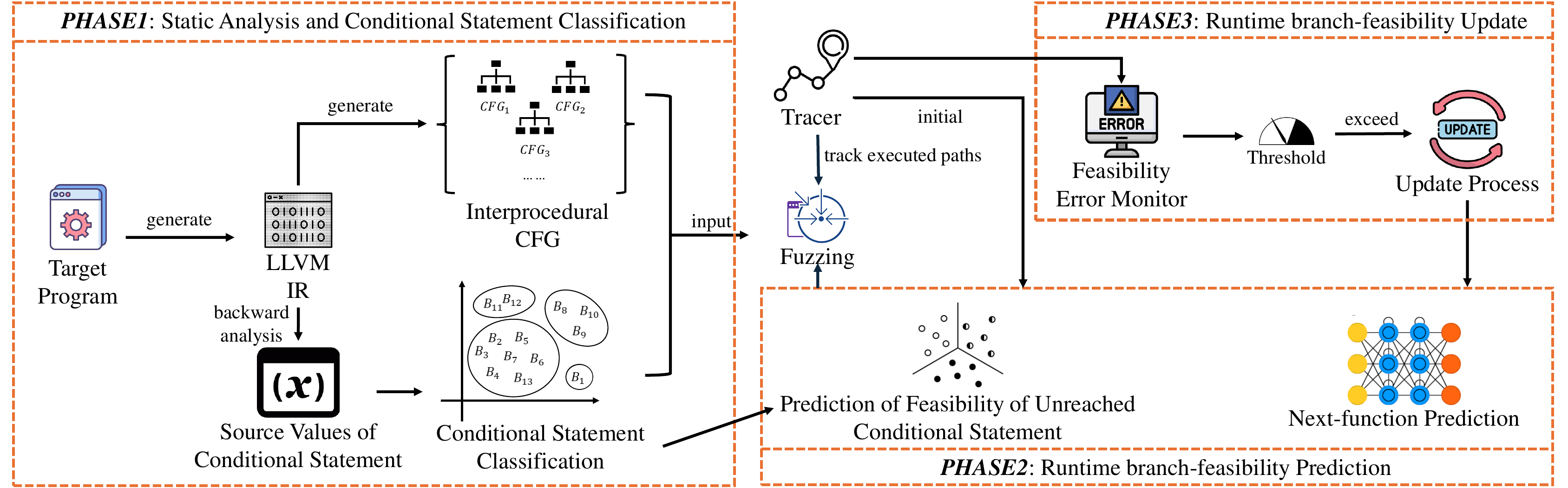}}
% \centerline{\includegraphics[width=1\textwidth]{graphs/whole.pdf}}
\caption{Overview of \sys.} 
\label{fig:overview}
\end{figure*}

The primary goal of directed fuzzing is to reach a specified target location in the program as efficiently as possible. 
Traditional distance-guided fuzzers such as AFLGo prioritize inputs based on static control-flow distance. 
However, these approaches fail to account for whether the paths are actually feasible under fuzzing conditions, often wasting resources on unreachable branches. 
\sys addresses this limitation by estimating the empirical \emph{feasibility} of branches and integrating this information into its guidance logic.

\autoref{fig:overview} illustrates the architecture of \sys, which is structured into three main phases: (1) \textit{static conditional-statement classification}, (2) \textit{runtime branch-feasibility prediction}, and (3) \textit{runtime branch-feasibility update}. 
These phases operate as a closed feedback loop to progressively guide fuzzing toward feasible paths leading to the target.

\PP{Phase 1: Static Analysis and Conditional Statement Classification}  
Given a target program, \sys first generates its LLVM IR and performs backward analysis to extract the source values of all conditional branch statements. These statements are grouped into semantically similar clusters based on attributes such as the role of the condition variable (e.g., pointer validity, return value checks), the comparison structure (e.g., null checks, threshold comparisons), and the usage context (e.g., memory allocation, I/O). An interprocedural control-flow graph (CFG) is also constructed to enable downstream distance computation. The resulting semantic clusters and CFG form the static inputs to the fuzzer.

\PP{Phase 2: Runtime Branch-Feasibility Prediction.}  
During fuzzing, \sys instruments and executes the program on input seeds. A tracer records which branches are exercised in the generated execution traces. 
For uncovered branches, \sys estimates feasibility using two mechanisms: (i) for conditional statements, feasibility is generalized from the observed behavior of semantically similar branches; 
and (ii) for indirect calls, \sys applies a next-function prediction model based on LSTM to estimate the probability of reaching a given call target. These feasibility scores are then embedded into a \textbf{feasibility-aware distance metric}, which replaces static CFG distance for seed ranking.

\PP{Phase 3: Runtime Branch-Feasibility Update.}  
As fuzzing proceeds, the system continuously monitors the prediction error between observed and expected branch behavior. If the error rate exceeds a predefined threshold, the \emph{Feasibility Error Monitor} triggers the \emph{Update Process}, which incrementally refines the feasibility scores—either by re-evaluating conditional clusters or updating the LSTM prediction model. This allows \sys to adapt in real time as new execution paths are uncovered.

\PP{Fuzzing Guidance Loop.}  
Throughout the fuzzing campaign, the feasibility-aware distance serves as the core signal for prioritizing seed inputs. Seeds that lie on paths with high feasibility and structural proximity to the target receive more mutation energy, as governed by the power schedule. This allocation strategy ensures that \sys systematically prioritizes promising paths that are both reachable and close—significantly improving its ability to reach deep or guarded target locations with minimal wasted effort.

\section{Design of \sys}

\sys is designed to accelerate directed fuzzing by prioritizing execution paths that are not only structurally close to the target but also empirically feasible. This section describes the key components of \sys and how it addresses two central challenges:  
(\textbf{C1}) how to predict the feasibility of unexplored branches, and  
(\textbf{C2}) how to refine these predictions at runtime as new execution traces become available.

To address (\textbf{C1}), \sys introduces two core techniques:  
(1) \emph{semantic clustering of conditional statements}, enabling generalization of feasibility across similar branches, and  
(2) \emph{next-function prediction} for estimating the reachability of indirect call targets.  
To handle (\textbf{C2}), \sys incorporates a lightweight runtime \emph{feasibility updating mechanism} that adapts predictions as coverage improves.

\subsection{Semantic Clustering of Conditional Statements}
\label{design:semantics}

The feasibility of a conditional branch is often governed not just by its syntax, but by the semantics of the data it evaluates. Our analysis and manual study in \autoref{sec:motivation} across real-world software reveal that the behavior of conditional branches is heavily influenced by the \emph{source} of the condition variable. To enable generalization of feasibility across uncovered code, \sys performs static semantic clustering of conditional statements based on their source characteristics.

We design a lightweight yet expressive taxonomy to capture the most common sources of conditional variables that affect fuzzing feasibility. Through a \emph{context-sensitive backward analysis} of LLVM IR across seven diverse open-source programs (including parsers, compressors, and system utilities), we identify five dominant categories that account for over 97\% of condition variable sources:

\noindent $\bullet$ \textbf{Function return values} — e.g., \texttt{malloc()}, \texttt{fopen()}, \texttt{xmlParse()}; often influence error-handling paths with high feasibility asymmetry (success is common, failure is rare).

\noindent $\bullet$ \textbf{Global variables} — e.g., initialization flags, configuration toggles; typically require specific program state setup, making certain branches hard to reach through input mutation alone.

\noindent $\bullet$ \textbf{Function arguments} — passed directly from input, commonly used in range, size, or validity checks.

\noindent $\bullet$ \textbf{Struct or object fields} — e.g., \texttt{req->hdr.len}; prevalent in input parsers or network handlers where correct structure fields govern execution paths.

\noindent $\bullet$ \textbf{Locals derived from input} — local variables calculated from input arguments or structs, such as \texttt{offset + size}; frequently used in nested logic and boundary guards.

These categories were chosen to strike a balance between semantic expressiveness, generalization capability, and implementation simplicity. While additional cases (e.g., environment variables or file descriptors) exist, they appear infrequently or are hard to analyze without dynamic context. Our taxonomy is extensible but empirically sufficient for broad applicability.

To extract these semantics, \sys performs a \emph{context-sensitive backward analysis} over LLVM IR to trace each condition variable to its origin. From the source, it extracts a semantic string that includes the variable type, variable name, and, if applicable, the associated function name or struct. These strings are then tokenized and embedded using a pre-trained Word2Vec model. Each conditional statement is represented as a concatenated embedding vector of its semantic tokens.

We cluster these semantic vectors using the DBSCAN algorithm. DBSCAN is chosen for its ability to find arbitrarily shaped clusters without requiring the number of clusters to be specified upfront—an important property for unsupervised classification over diverse programs. Additionally, DBSCAN is robust to noise and outliers, making it well-suited for handling imperfect static data.

The resulting clusters group conditional statements that are semantically similar in terms of how they guard program logic. These clusters form the foundation for \sys’s feasibility generalization mechanism at runtime: when an uncovered branch belongs to a known cluster, \sys estimates its feasibility based on the aggregated behavior of covered branches in the same group. This approach allows \sys to extend feasibility knowledge to parts of the program that have not yet been executed, improving both guidance quality and fuzzing efficiency (detail in \autoref{sec:FPred}).

\subsection{Feasibility Prediction}
\label{sec:FPred}

At runtime, \sys estimates the feasibility of both covered and uncovered branches using two distinct strategies:

\PP{Conditional Branches Feasibility Prediction}  
For conditional statements, \sys collects execution traces and aggregates feasibility statistics within each semantic cluster. Specifically, it computes the average feasibility of all covered branches in a cluster and assigns this value to its uncovered members. This reflects our hypothesis that semantically similar conditionals—such as those checking the return value of memory allocation functions—are likely to exhibit similar branch-taking behavior. When a cluster contains no covered branches (e.g., early in fuzzing), we assign it the feasibility of the \emph{nearest cluster} in vector space, measured by centroid distance. This fallback is motivated by the observation that clusters close in semantic embedding space often share similar operational intent. Compared to alternatives such as using a global average or uniform default, nearest-cluster inference preserves semantic locality and avoids diluting the guidance signal with unrelated behavior.

\PP{Indirect Call Feasibility Prediction}
\label{impl:indirect}
Unlike prior work such as CALLEE~\cite{b66}, which attempts to dynamically resolve indirect call targets during execution, \sys does not predict \emph{which} function will be invoked at runtime. Instead, it statically enumerates all possible indirect call targets and predicts the \emph{feasibility} of exercising each edge—i.e., the feasibility (likelihood) that a given target will actually be taken under fuzzing.

To construct the candidate set of indirect call targets, \sys integrates Multi-Layer Type Analysis (MLTA)~\cite{b34}. MLTA is specifically chosen because it guarantees \textbf{zero false negatives}, ensuring that no legitimate targets are omitted. Compared to conventional pointer or alias analysis, MLTA provides a sound and efficient overapproximation of the indirect call graph. This results in a statically computed call graph where each indirect call site is mapped to a complete set of feasible target functions.

The core challenge, then, is not target discovery—but estimating \emph{how likely} each of these statically identified targets is to be reached during fuzzing. This task presents three distinct difficulties: (1) each indirect call may dispatch to multiple targets; (2) the static context surrounding each indirect call target is often identical—since all targets are invoked from the same call site—making it ineffective to distinguish their feasibility using static features. As a result, we cannot apply the same semantic clustering strategy used for conditional statements, which relies on variation in static context to generalize feasibility.; and (3) early fuzzing coverage is sparse, limiting the training data available for target-specific learning.

To address this, \sys introduces a technique we call \emph{next-function feasibility prediction}, inspired by next-word prediction in natural language processing. Execution traces are collected and tokenized as sequences of function calls, where the goal is to predict the feasibility of each indirect call edge based on the preceding call chain. For example, given a trace like \texttt{main → parse → dispatch → ?}, the model learns to assign probabilities to candidate targets observed at the final indirect call site.

We implement this prediction using a single-direction Long Short-Term Memory (LSTM) network. LSTM is selected over alternatives such as Bi-LSTM or Transformer-based models for several practical reasons. First, the function-call traces are inherently directional—only the forward context is relevant at the time of the indirect call. Bi-LSTM introduces backward dependencies that are semantically meaningless in this context. Second, LSTMs are more computationally efficient for the short sequences we typically observe (10–200 tokens), with $O(nd^2)$ complexity compared to $O(n^2d)$ for Transformer-based models. Finally, LSTMs support online and incremental learning, making them well-suited for the evolving nature of fuzzing where traces are collected continuously and model updates must be lightweight.

To overcome the initial lack of indirect call training data, \sys employs \emph{predominating functions}—functions that frequently appear in traces leading to indirect calls—as proxy labels. These allow early-stage training while preserving the directionality and structure of traces. As coverage improves and actual indirect call targets are reached, the model transitions to learning target-specific feasibility scores. The LSTM outputs a softmax distribution over candidate targets for each call site; these scores are used as feasibility estimates for edge weighting in the call graph.

Model updates are coordinated by \sys’s runtime error monitoring framework (detail in \autoref{sec:DT}). When the prediction error exceeds a threshold $\theta_{IC}$, the model is retrained incrementally using newly collected traces. The LSTM module is implemented in PyTorch and integrated with the main fuzzing loop via lightweight Inter-Process Communication (IPC), allowing predictions and updates to run asynchronously without degrading fuzzing throughput.

\subsection{Runtime Feasibility Update}
\label{sec:DT}

As fuzzing progresses, new execution traces continually reveal branches and indirect call targets that were previously uncovered. To ensure guidance remains accurate and effective, \sys employs a dynamic runtime updating mechanism that refines feasibility predictions based on observed behavior. Without such adaptivity, static feasibility estimates risk becoming outdated or misleading—particularly in programs with highly skewed or input-sensitive control flow.

\sys's update mechanism is driven by an error-monitoring system that evaluates the divergence between prior predictions and newly observed behavior. When prediction errors exceed predefined thresholds, the system triggers selective updates to restore model accuracy and improve future guidance. We separately monitor prediction error for (1) conditional statement clusters and (2) indirect call targets, using distinct but harmonized metrics.

\PP{Conditional Branch Monitoring}
For conditional statements, \sys tracks the prediction error for each semantic cluster. We define $Err_{CSC}$ as the relative change in feasibility between two update cycles, calculated as 

\begin{equation}\label{eq:err_csc}
    Err_{CSC} = \frac{|P' - P|}{P}
\end{equation}

where $P$ and $P'$ are the old and new feasibility estimates for the cluster. If a cluster's error exceeds a threshold $\theta_{CSC}$, it is marked as unstable. We then compute the overall group error rate, 

\begin{equation}\label{eq:err_g}
    Err_G = \frac{N'}{N}
\end{equation}

where $N'$ is the number of unstable clusters and $N$ is the total number of clusters.

When $Err_G>\theta_G$, \sys initiates an update: the average feasibility for each cluster is recalculated, and feasibility-aware distances are recomputed accordingly. This ensures that fuzzing energy remains focused on truly promising paths rather than stale estimations.

\PP{Indirect Call Monitoring.}
For indirect-call targets, \sys leverages a prediction model trained on function-call sequences. We monitor the model's accuracy $\theta_{acc}$ before and after each training round. The prior error rate is defined as 

\begin{equation}\label{eq:err_g1}
    Err_{IC} = 1 - \theta_{acc}
\end{equation}

and the new error rate as $Err'_{IC}$. When 

\begin{equation}\label{eq:err_g2}
    |Err_{IC} - Err'_{IC}| > \theta_{IC}
\end{equation}

we trigger model retraining using the most recent execution traces.

\PP{Update Execution}
Updates occur after each fuzzing cycle (as defined by AFL), but are only executed if thresholds are exceeded. Once triggered, the following steps occur:
(1) feasibility scores for relevant edges are updated,  
(2) the dynamic distance table is refreshed with new weights, and  
(3) guidance rankings for seeds are recalculated.  
This ensures that updates remain lightweight and do not interfere with fuzzing throughput.

\PP{Threshold Selection.}
Optimal thresholds are determined empirically using the control variate method across nine real-world programs. We varied one threshold at a time and observed fuzzing effectiveness and update cost. Results indicate that setting $\theta_{CSC}=10\%$ and $\theta_G=3\%$ provides a good trade-off between update frequency and prediction accuracy. Lower thresholds resulted in excessive updates and performance overhead; higher thresholds led to outdated guidance. Users may adjust these parameters for domain-specific tuning (detail in \autoref{sec:impact of update}).

\PP{Dynamic Distance Table.}
To efficiently reflect updates in fuzzing guidance, \sys replaces AFLGo's static distance instrumentation with a dynamic distance table. This table stores basic block and function edge weights in memory and is updated in-place when feasibility changes. This eliminates the need to recompile the target binary and significantly reduces guidance update latency.

\PP{Incremental Learning for Indirect Call Feasibility.}
To support timely and efficient updates to the next-function prediction model, \sys employs incremental learning. New traces are used to retrain the LSTM model without discarding prior knowledge. This reduces retraining time and ensures the model adapts gracefully as new indirect call targets are discovered. The model is implemented in PyTorch and integrated via IPC with the fuzzer, allowing asynchronous prediction and retraining without largely impacting fuzzing speed.

\section{Implementation}

\sys is implemented atop LLVM and AFLGo, with key components written in C++, Python, and PyTorch. The system consists of four primary modules: (1) static analysis and interprocedural graph construction, (2) normalization and classification of conditional statements, (3) feasibility-aware distance calculation, and (4) a dynamic runtime updating mechanism. Below, we detail the key implementation considerations for each module.

\subsection{Constructing the Interprocedural CFG}

Accurate interprocedural control-flow representation is critical for directed fuzzing, especially when indirect calls and nested function interactions are involved. \sys constructs an interprocedural control-flow graph (Inter-CFG) by combining intra-procedural control-flow graphs (CFGs) with a statically resolved call graph (CG).

To resolve indirect calls, we integrate Multi-Layer Type Analysis (MLTA)~\cite{b34}, which provides a sound and efficient overapproximation with \textbf{zero false negatives}. This guarantees that all feasible indirect call targets are preserved. Compared to alias or pointer analysis, MLTA strikes an effective balance between precision and coverage. We integrate MLTA with LLVM’s IR to build the Inter-CFG, which is then used for distance computations and feasibility-aware guidance throughout fuzzing.

\subsection{Normalization of Conditional Statements}
\label{sec:impl-normalization}

Prior to semantic clustering and feasibility prediction, conditional statements must be normalized to ensure consistency. Syntactically different but semantically equivalent conditions can lead to divergent feasibility estimates if not handled correctly. For example, consider \texttt{if (!x)} vs. \texttt{if (x == NULL)} — both check for nullity, but without normalization, they may be assigned different feasibility labels.

To address this, \sys treats conditions using \texttt{==}, \texttt{>}, and \texttt{<} as standard forms. Conditions with \texttt{!=}, \texttt{!}, \texttt{<=}, and \texttt{>=} are transformed by inverting the branch feasibility assignment. For compound conditionals using \texttt{\&\&} or \texttt{||}, LLVM IR naturally decomposes them into atomic conditions, which \sys handles as separate single-branch checks during analysis and modeling.

\subsection{Feasibility-Aware Distance Calculation}

\sys extends AFLGo’s distance computation methodology by modifying how edge weights are assigned. In AFLGo, all edges are weighted uniformly (typically with weight 1), which leads to feasibility-unaware guidance. In contrast, \sys dynamically assigns edge weights based on empirical or predicted feasibility, with lower weights indicating higher likelihood of execution.

\paragraph{Basic Block Level.}
At the intra-procedural level, each edge between basic blocks is weighted using the reciprocal of its feasibility: 

\begin{equation}\label{eq:bb_weight}
    W_{\text{BB\_edges}} = \frac{1}{P_{\text{BB}}}
\end{equation}

where $P_{BB}$ is the empirical or predicted feasibility of the branch. High-feasibility edges thus have lower weights, biasing the distance metric toward paths that are both short and likely to be executed.

\paragraph{Function Level.}
Function-level edge weights are more nuanced, as function calls may be gated by one or more conditional branches. To account for this, we introduce the concept of a \emph{Conditional Statement Chain (CSC)}—a series of conditional statements that dominate a particular function call site. We statically identify CSCs during control-flow analysis.

For each function call, we compute its feasibility $P_F$ based on the enclosing CSCs:

\begin{equation}\label{eq:pd}
    P_F = 
    \begin{cases}
    1 & \text{if } FC \notin \text{any CSC}, \\
    \max\limits_{i} \left( \prod\limits_{j=1}^n P_{ij} \right) & \text{if } FC \in \{CSC_i\}, \\
    \end{cases}
\end{equation}

where $P_{ij}$ is the feasibility of the $j$-th condition in CSC $i$. We take the maximum across CSCs to favor the most permissive path. The function-level edge weight is then computed as 

\begin{equation}\label{eq:pd1}
    W_{F\_edges} = \frac{1}{P_F}
\end{equation}
, ensuring that paths through easily satisfiable conditions are prioritized.

\subsection{Dynamic Distance Table}

To support runtime feasibility updates, \sys replaces AFLGo’s static distance instrumentation with a dynamic in-memory distance table. This table stores weights for both basic block and function edges and is initialized at program startup. When new feasibility estimates are computed (via trace analysis or model updates), only the relevant entries in the table are updated. This avoids recompilation and allows the fuzzer to respond to new information in near real-time. 
% Further analysis of update latency and effectiveness is provided in \autoref{sec:new-tech}~\WB{Double check}.

\subsection{Runtime Prediction and Incremental Model Updates}

For indirect-call edges, \sys predicts edge feasibility using an LSTM-based model trained on dynamic function-call traces. The model is implemented in PyTorch and runs as a standalone module, interfacing with the fuzzer through a dedicated tracker. This tracker monitors execution coverage, enabling real-time updates of feasibility information.

During fuzzing, the fuzzer extracts the current function-call sequence and sends it to the prediction module. The LSTM processes this sequence and returns a softmax vector over the statically collected indirect call targets (from MLTA). Each entry in the vector represents the predicted feasibility of reaching a specific target. These scores are then translated into edge weights using the reciprocal transformation described in Equation~\ref{eq:bb_weight}, and used to update the feasibility-aware distance table in memory.

To ensure responsiveness, the LSTM model is trained incrementally. New execution traces are batched and used to update the model from its last checkpoint, avoiding costly retraining. Updates are triggered by \sys’s error monitoring system (\autoref{sec:DT}) whenever prediction error exceeds a defined threshold. Because both training and inference operate asynchronously from the main fuzzing loop, they introduce no measurable slowdown in fuzzing throughput.

This integration of on-the-fly model refinement with in-place edge weight updates allows \sys to continuously adapt its guidance strategy as new execution paths are discovered. It ensures that feasibility predictions remain accurate throughout the fuzzing campaign and that distance metrics reflect the evolving program state.

\section{Evaluation}
\label{sec:eval}

Directed fuzzing aims to reach and reproduce user-specified target locations—such as patched code or known vulnerabilities—as efficiently as possible. To evaluate the effectiveness of \sys in fulfilling this objective, we conduct a large-scale experimental study. This section answers the following research questions:

\noindent \textbf{RQ1:} Does \sys reach predefined target sites more effectively than state-of-the-art directed fuzzers?

\noindent \textbf{RQ2:} Is \sys more effective in reproducing known bugs than competing approaches?

\noindent \textbf{RQ3:} How much do \sys’s individual design components contribute to its overall effectiveness?

\subsection{Evaluation Setup}

All experiments are performed on a virtual machine running Ubuntu 18.04 with 4GB RAM and 50GB disk space, hosted on a 64-bit Windows 11 Pro system with an Intel Core i9-12900K @ 3.20GHz, 128GB of RAM, and a 24GB Nvidia RTX 3090Ti GPU. The fuzzing environment aligns with prior fuzzing research such as Healer~\cite{b63}.

\subsubsection{Evaluation Baselines}

We compare \sys against a diverse set of baseline fuzzers:

\noindent $\bullet$ \textbf{AFLGo}~\cite{b3} pioneered the concept of distance-guided directed fuzzing and remains a standard benchmark.  

\noindent $\bullet$ \textbf{AFLGo-ICall} is our enhanced version that extends AFLGo to support indirect call edges in the call graph.

\noindent $\bullet$ \textbf{WindRanger}~\cite{b54}, which steers fuzzing toward deviation basic blocks;  

\noindent $\bullet$ \textbf{BEACON}~\cite{b61}, which employs path pruning to exclude infeasible branches; 

\noindent $\bullet$ \textbf{SelectFuzz}~\cite{b42}, which selectively explores code regions relevant to the target site.

\noindent $\bullet$ \textbf{DeepGo}~\cite{b70}, which claims to be open source, but we were unable to obtain or reproduce their source code at the time of writing. As such, we will compare with them based on the result provided in their paper.

\subsubsection{Evaluation Metrics and Methodology}
\label{sec:metrics}

We follow prior work~\cite{b2,b3,b11,b30,b54} in adopting two key \textbf{metrics}:

\noindent $\bullet$ \textbf{Time-to-Target (TTT)}: Time required for a fuzzer to first \emph{reach} a predefined target location in the binary.  

\noindent $\bullet$ \textbf{Time-to-Exposure (TTE)}: Time required to first \emph{trigger} a known bug associated with the target.

\PP{Experimental Methodology.}  
For each target site, we run each fuzzer 10 times with a fixed 24-hour time budget per run to reduce randomness and ensure statistical robustness. If a fuzzer fails to reach the target or expose the bug in all runs, we mark the result as \textit{Timeout (T.O.)}.
Meanwhile, for some target programs, the baseline fuzzer cannot compile, we will mark as \textbf{N/A}, which will exclude from the comparison of the performance.

To assess statistical significance, we apply the Mann–Whitney U test ($p$-value)~\cite{b55}. We consider results statistically significant when $p < 0.05$, allowing us to reject the null hypothesis. Additionally, we use the Vargha–Delaney $\hat{A}_{12}$ effect size~\cite{b56} to quantify whether one technique outperforms another. A value of $\hat{A}_{12} > 0.71$ is interpreted as a large performance difference.

\PP{Data sampling for conditional statement classification verification}   
To validate the consistency of feasibility across semantically similar conditional statements, we perform manual analysis on a representative benchmark dataset collected by us. Following practices from recent program analysis and fuzzing studies~\cite{bavishi2019phoenix, fu2021cpscan, gens2018k, lu2019detecting, vassallo2020developers, wu2024gnnic}, which typically analyze 40–400 samples, we conservatively select 1,000 conditional statements for inspection based on pre-defined rules (see \autoref{sec:motivation}). All sampled conditionals are cross-validated by three developers with more than five years of C/C++ experience. This process ensures that our classification and feasibility evaluations are unbiased, statistically grounded, and reproducible. According to empirical sampling theory~\cite{conroy2015sample}, this sample size yields a margin of error of approximately ±5\%, offering strong statistical confidence for our consistency checks.

\subsubsection{Benchmark Dataset}
\label{s:dataset}

To ensure fairness, reproducibility, and comparability with prior work, we evaluate \sys using the same benchmark datasets adopted by AFLGo~\cite{b3} and widely used in recent directed fuzzing research~\cite{b2, b11, b54, b61, b70, b42}.

\noindent $\bullet$ \textbf{UniBench}~\cite{li2021unifuzz} contains real-world programs from diverse domains  along with curated seed corpora. This dataset has been adopted by several recent directed fuzzers, including WindRanger~\cite{b54} and DeepGo~\cite{b70}. Although DeepGo's implementation is not publicly available, we use the same target programs and positions to support a theoretical comparison under RQ1.

\noindent $\bullet$ \textbf{AFLGo Test Suite (ATS)}\footnote{\url{https://github.com/aflgo/aflgo/tree/master/examples}} was introduced in AFLGo’s original work~\cite{b3} to measure a fuzzer’s ability to reach known vulnerabilities. It includes ground truth crash sites and target locations, making it well-suited for evaluating bug reproduction capabilities. To answer RQ2, we adopt the AFLGo test suite as a benchmark to assess how effectively each fuzzer can expose known vulnerabilities under time-constrained conditions.

By using these two well-established benchmarks, we ensure that our results are directly comparable to those of prior work, and that both target-reaching and bug-triggering aspects of directed fuzzing are thoroughly evaluated.

\subsection{Effectiveness on Reaching Target Sites}
\label{sec:TTT}

To evaluate the effectiveness of \sys in reaching predefined target sites, we tested 20 programs from the UniBench benchmark suite, covering a total of 80 unique target locations. For each target, we measured the Time-to-Target (TTT), defined as the time required for a fuzzer to reach the designated code location. Each experiment was executed with a 24-hour timeout threshold per run, and results were averaged over 10 independent runs. Full results are summarized in \autoref{tb:TTT} (Due to font restriction, full results appear in the Appendix).

\sys successfully reached 77 out of 80 target sites, substantially outperforming all baselines: AFLGo (7/80), BEACON (10/80), WindRanger (15/80), SelectFuzz (27/80), and AFLGo-ICall (29/80). Furthermore, \sys achieved the shortest average TTT on 71 out of 80 targets, demonstrating its consistent superiority across a wide range of program types and target patterns.

In terms of speed, \sys delivers substantial improvements, achieving 3.76$\times$, 2.57$\times$, 3.30$\times$, and 2.52$\times$ average speedup in TTT compared to AFLGo, BEACON, WindRanger, and SelectFuzz, respectively. When compared to AFLGo-ICall, \sys achieves a 2.86$\times$ speedup. Although DeepGo~\cite{b70} is not publicly available for direct comparison, we approximate its relative performance based on the TTT results it reported. When normalized via AFLGo, BEACON, and WindRanger as anchor baselines, \sys demonstrates estimated TTT improvements of 1.17$\times$, 1.49$\times$, and 1.82$\times$ over DeepGo, respectively.

To ensure statistical rigor, we conducted both the Mann–Whitney U test and the Vargha–Delaney effect size test. All $p-values$ were below 0.05, indicating statistically significant improvements. The mean $\hat{A}_{12}$ values of \sys versus AFLGo, BEACON, WindRanger, SelectFuzz, and AFLGo-ICall were 0.78, 0.81, 0.86, 0.82, and 0.79, respectively—reflecting large effect sizes according to standard interpretation~\cite{b56}.

\textbf{Conclusion:} Across 80 diverse targets, \sys consistently reaches more targets and does so significantly faster than all evaluated baselines, validating its advantage in target-oriented fuzzing.

\subsection{Effectiveness of Vulnerability Reproduction}

\begin{table*}
    \centering
        \centering
        \caption{The Time-to-Exposure Results on Programs from ATS}
        \resizebox{\linewidth}{!}{
            \begin{tabular}{ccccccccccccc}
\toprule
       \textbf{Program} &     \textbf{CVE-ID} &         \textbf{AFLGo} &        \textbf{BEACON} &       \textbf{WindRanger} &    \textbf{SelectFuzz} &       \textbf{AFLGo-ICall} &   \textbf{AFLGopher-C} &   \textbf{AFLGopher-N} & \textbf{AFLGopher-U} & \textbf{AFLGopher} \\
\toprule
\multirow{7}*{binutils2.26} &  2016-4487 &         0.16h &         0.05h &         0.14h &         0.18h &         0.17h &         0.35h &         0.42h &       0.33h &     0.33h \\
             &  2016-4488 &          0.4h &         3.38h &         0.26h &         0.53h &         0.39h &         1.19h &         1.21h &       0.58h &     0.57h \\
             &  2016-4489 &         0.27h &         0.31h &         0.42h &          0.3h &          0.3h &         0.44h &         0.75h &       0.73h &     0.72h \\
             &  2016-4490 &          0.1h &         0.17h &         0.28h &         0.39h &         0.17h &         0.33h &         0.47h &        0.3h &      0.3h \\
             &  2016-4491 & \textit{T.O.} & \textit{T.O.} & \textit{T.O.} &        18.26h &        23.24h &         8.24h & \textit{T.O.} &       7.95h &     4.11h \\
             &  2016-4492 &          0.9h &         0.79h &         0.16h &         1.89h &         0.92h &         1.11h &         1.64h &       1.05h &     0.45h \\
             &  2016-6131 & \textit{T.O.} & \textit{T.O.} &         21.4h &        19.25h & \textit{T.O.} &        12.46h & \textit{T.O.} &      11.55h &     4.25h \\
\midrule
\multirow{4}*{libming4.48} &  2018-8807 & \textit{T.O.} &        11.86h &        13.88h &        18.18h &        21.39h &         16.4h &        22.42h &       11.3h &      4.4h \\
             &  2018-8962 &        19.36h &         8.12h &         13.2h &        14.42h &        17.87h &         13.9h &        12.21h &       7.73h &     3.49h \\
             & 2018-11095 & \textit{T.O.} & \textit{T.O.} & \textit{T.O.} & \textit{T.O.} & \textit{T.O.} & \textit{T.O.} & \textit{T.O.} &      10.68h &     4.78h \\
             & 2018-11225 & \textit{T.O.} &        17.34h & \textit{T.O.} &        20.41h &        22.99h &        19.33h & \textit{T.O.} &      16.57h &     4.72h \\
\midrule
\multirow{3}*{LibPNG1.5.1} &  2011-2501 &         0.79h &           N/A &         0.73h &         1.84h &         1.24h &         1.95h &         1.67h &       1.32h &     0.36h \\
             &  2011-3328 &          7.7h &           N/A &         4.32h &         7.49h &         7.64h &         4.08h &         3.61h &        2.8h &     1.35h \\
             &  2015-8540 &         0.06h &           N/A &         0.11h &         1.97h &         0.11h &          0.4h &          0.4h &        0.4h &      0.4h \\
\midrule
\multirow{4}*{xmllint2.9.4} &  2017-9047 & \textit{T.O.} & \textit{T.O.} & \textit{T.O.} &        18.39h & \textit{T.O.} &         9.12h &        12.48h &       8.74h &     4.27h \\
             &  2017-9048 & \textit{T.O.} & \textit{T.O.} & \textit{T.O.} &         2.51h &         2.27h &         3.05h & \textit{T.O.} &       1.73h &     0.57h \\
             &  2017-9049 & \textit{T.O.} & \textit{T.O.} & \textit{T.O.} &        21.41h & \textit{T.O.} &        17.98h & \textit{T.O.} &       6.93h &      4.4h \\
             &  2017-9050 & \textit{T.O.} & \textit{T.O.} & \textit{T.O.} &        21.09h & \textit{T.O.} &        19.99h & \textit{T.O.} &      13.19h &     4.17h \\
\midrule
\multirow{2}*{Lrzip0.631} &  2017-8846 & \textit{T.O.} &        11.05h &        17.74h &         21.1h & \textit{T.O.} &        12.13h & \textit{T.O.} &      11.64h &     4.49h \\
             & 2018-11496 &        22.46h &          9.3h &        19.11h &        22.31h &        21.91h &        11.99h &         9.62h &       7.93h &     4.01h \\
\toprule
\multicolumn{2}{c}{\textbf{\textit{speedup}}} & \multicolumn{1}{c}{\textbf{5.60x}} & \multicolumn{1}{c}{\textbf{5.20x}} & \multicolumn{1}{c}{\textbf{4.98x}} & \multicolumn{1}{c}{\textbf{4.52x}} & \multicolumn{1}{c}{\textbf{5.07x}} & \multicolumn{1}{c}{\textbf{3.42x}} & \multicolumn{1}{c}{\textbf{4.97x}} & \multicolumn{1}{c}{\textbf{2.37x}} & \multicolumn{1}{c}{\textbf{-}}\\
\multicolumn{2}{c}{\textbf{mean \textit{p-values}}} & \multicolumn{1}{c}{\textbf{0.03}} & \multicolumn{1}{c}{\textbf{0.002}} & \multicolumn{1}{c}{\textbf{0.004}} & \multicolumn{1}{c}{\textbf{0.003}} & \multicolumn{1}{c}{\textbf{0.006}} & \multicolumn{1}{c}{\textbf{0.014}} & \multicolumn{1}{c}{\textbf{0.011}} & \multicolumn{1}{c}{\textbf{0.017}}    & \multicolumn{1}{c}{\textbf{-}} \\
\multicolumn{2}{c}{\textbf{mean $\hat A_{12}$}} & \multicolumn{1}{c}{\textbf{0.76}} & \multicolumn{1}{c}{\textbf{0.83}} & \multicolumn{1}{c}{\textbf{0.89}} & \multicolumn{1}{c}{\textbf{0.80}} & \multicolumn{1}{c}{\textbf{0.81}} & \multicolumn{1}{c}{\textbf{0.79}} & \multicolumn{1}{c}{\textbf{0.86}} & \multicolumn{1}{c}{\textbf{0.88}} & \multicolumn{1}{c}{\textbf{-}}\\
\bottomrule
\end{tabular} % Your first table data
        }
        \label{tb:TTE}
\end{table*}

To evaluate how effectively \sys exposes known vulnerabilities, we follow the same setup as BEACON~\cite{b61}, WindRanger~\cite{b54}, and DeepGo~\cite{b70} by using the AFLGo test suite, which defines 20 real-world CVE-labeled vulnerabilities as target sites. These target locations and the corresponding Time-to-Exposure (TTE) results are presented in \autoref{tb:TTE}.

Among all evaluated fuzzers, \sys uncovered the most vulnerabilities (20/20), outperforming AFLGo (11), BEACON (10), WindRanger (13), SelectFuzz (19), and AFLGo-ICall (16). Moreover, \sys achieved the shortest TTE on 15 out of 20 targets, demonstrating its consistent efficiency across a diverse set of real-world programs.

In terms of mean TTE, \sys delivers substantial speedups over all baselines: 5.60$\times$ over AFLGo, 5.20$\times$ over BEACON, 4.98× over WindRanger, 4.52$\times$ over SelectFuzz, and 5.07$\times$ over AFLGo-ICall. All differences are statistically significant ($p < 0.05$), and the corresponding mean $\hat{A}_{12}$ effect sizes are 0.76, 0.83, 0.89, 0.80, and 0.81, respectively—indicating large performance improvements~\cite{b56}.

Although DeepGo is not publicly available for direct evaluation, we estimate relative performance based on its published TTE results. Using AFLGo, BEACON, and WindRanger as normalization anchors, we estimate that \sys achieves 2.14$\times$, 1.57$\times$, and 2.05$\times$ speedups over DeepGo, respectively. These indirect comparisons further suggest that \sys advances the state of the art in vulnerability exposure.

\textbf{Conclusion:} \sys consistently exposes known vulnerabilities faster than all evaluated baselines, and shows promising improvements over reported state-of-the-art systems, validating its effectiveness in vulnerability-triggering scenarios under realistic fuzzing constraints.

\subsection{Effectiveness and Accuracy of Conditional Statement Classification (CSC)}
\label{sec:csc-eval}

To assess the contribution of our conditional statement classification (CSC) module, we evaluate its impact on both Time-to-Target (TTT) and Time-to-Exposure (TTE) across two benchmark suites: UniBench and the AFLGo Test Suite (ATS). For this purpose, we isolate the CSC component in \sys, referred to as \sys-C, and compare its performance against all baseline fuzzers and the full version of \sys.

As shown in \autoref{tb:TTT} and \autoref{tb:TTE}, \sys-C achieves consistent improvements over AFLGo and other baselines, though it does not match the full capabilities of \sys. On average, \sys-C achieves speedups of 1.52$\times$, 1.05$\times$, 1.34$\times$, 1.03$\times$, and 1.16$\times$ over AFLGo, BEACON, WindRanger, SelectFuzz, and AFLGo-ICall on TTT. On TTE, \sys-C delivers even more significant gains—1.64$\times$, 1.53$\times$, 1.46$\times$, 1.33$\times$, and 1.49$\times$ against the same fuzzers, respectively. These results demonstrate that CSC alone contributes meaningfully to the overall efficiency of \sys.

\PP{Clustering Accuracy}
To validate the accuracy of CSC, we perform both internal and external evaluations.

\noindent \emph{(1) Internal validation.}  
We manually labeled 1,000 conditional statements drawn from our benchmark dataset (see \autoref{sec:motivation} and \autoref{sec:metrics}) and use the Adjusted Rand Index (ARI)~\cite{steinley2004properties} to compare our semantic clusters against ground truth labels. Our clustering yields an ARI score of 0.91, indicating near-perfect alignment. In contrast, random clusterings on the same dataset average an ARI of 0.29, underscoring the effectiveness of our clustering method.

\noindent \emph{(2) External validation.}  
To further assess clustering quality, we use the Silhouette Coefficient (SC)~\cite{b65}, a standard unsupervised clustering metric that evaluates cohesion and separation. SC values range from -1 to 1, where values close to 1 indicate well-separated clusters, values near 0 suggest overlapping boundaries, and negative values indicate likely misclassifications. Across all semantic clusters generated by CSC, we observe an average SC of 0.82, with all cluster values exceeding 0.8—strong evidence that our clusters are both compact and well-separated.

Together, these quantitative results confirm that our conditional statement classification is not only semantically meaningful but also empirically precise, contributing directly to \sys's performance advantage.

\subsection{Effectiveness and Accuracy of Next Function Prediction (NFP)}
\label{sec:nfp-eval}

To assess the contribution of the Next Function Prediction (NFP) component, we evaluate its impact on both Time-to-Target (TTT) and Time-to-Exposure (TTE). For this analysis, we isolate NFP from \sys—referred to as \sys-N—and compare its performance to all baseline fuzzers as well as the full version of \sys.

As shown in \autoref{tb:TTT}, \sys-N achieves speedups of 1.40$\times$, 0.96$\times$, 1.22$\times$, 0.94$\times$, and 1.07$\times$ over AFLGo, BEACON, WindRanger, SelectFuzz, and AFLGo-ICall, respectively, in terms of average TTT. Similarly, in \autoref{tb:TTE}, \sys-N achieves TTE speedups of 1.13$\times$, 1.05$\times$, 1.01$\times$, 0.91$\times$, and 1.03$\times$ against the same set of fuzzers.

While NFP yields improvements in several scenarios, its standalone impact is less pronounced than that of CSC. This is largely attributable to the distribution of indirect calls across the evaluated programs. For example, \textsc{libming} contains over 300 indirect call sites, providing ample opportunity for NFP to optimize guidance. In such programs, NFP significantly enhances the feasibility estimation of indirect-call edges. However, in programs with relatively few indirect calls, the contribution of NFP is minimal. In contrast, conditional statements are pervasive in all programs, making CSC generally more influential when used in isolation.

These observations underscore that NFP and CSC are complementary: each addresses a different dimension of path feasibility. Their combined use is critical to \sys’s overall effectiveness.

\PP{Prediction Accuracy.}  
To evaluate the accuracy of the NFP model, we define precision using the following formula:

\[
\text{Accuracy} = 1 - \frac{1}{n} \sum_{i=1}^{n} \left| F_{p_i} - F_{t_i} \right|
\]

where \( n \) is the total number of indirect-call edges, \( F_{p_i} \) is the predicted feasibility score, and \( F_{t_i} \) is the ground truth feasibility from execution traces. Using this metric, NFP achieves a high average prediction accuracy of 95.11\%, indicating that the model is able to estimate feasibility with strong alignment to observed behavior.

\subsection{Effectiveness and Accuracy of Runtime Update Process}
\label{sec:impact of update}

To assess the contribution of the runtime update process, we evaluate its impact on both Time-to-Target (TTT) and Time-to-Exposure (TTE). For this analysis, we isolate this process from \sys—referred to as \sys-U—and compare its performance to all baseline fuzzers as well as the full version of \sys.

As shown in \autoref{tb:TTT}, \sys-U achieves speedups of 1.68$\times$, 1.15$\times$, 1.48$\times$, 1.14$\times$, and 1.29$\times$ over AFLGo, BEACON, WindRanger, SelectFuzz, and AFLGo-ICall, respectively, in terms of average TTT. Similarly, in \autoref{tb:TTE}, \sys-N achieves TTE speedups of 2.38$\times$, 2.20$\times$, 2.11$\times$, 1.91$\times$, and 2.14$\times$ against the same set of fuzzers.

These results corroborate that combining CSC and NFP outperforms either component alone. Although \sys-U improves over existing baselines, it still underperforms the full \sys because, without the update process, it cannot incorporate feasibility information from new traces during fuzzing.

\PP{Parameter Sensitivity: Thresholds for Triggering Updates}
The error thresholds $\theta_{CSC}$ and $\theta_G$ serve as key hyperparameters for \sys’s runtime update mechanism, determining when to retrain the prediction model and refresh feasibility-aware distances. To identify effective values, we apply a control variate methodology based on average Time-to-Target (TTT), varying one parameter while keeping the other fixed to isolate its individual impact.

\begin{figure}[htbp]
\centering
%\linewidth
\resizebox{0.48\textwidth}{!}{
\includegraphics[scale=1]{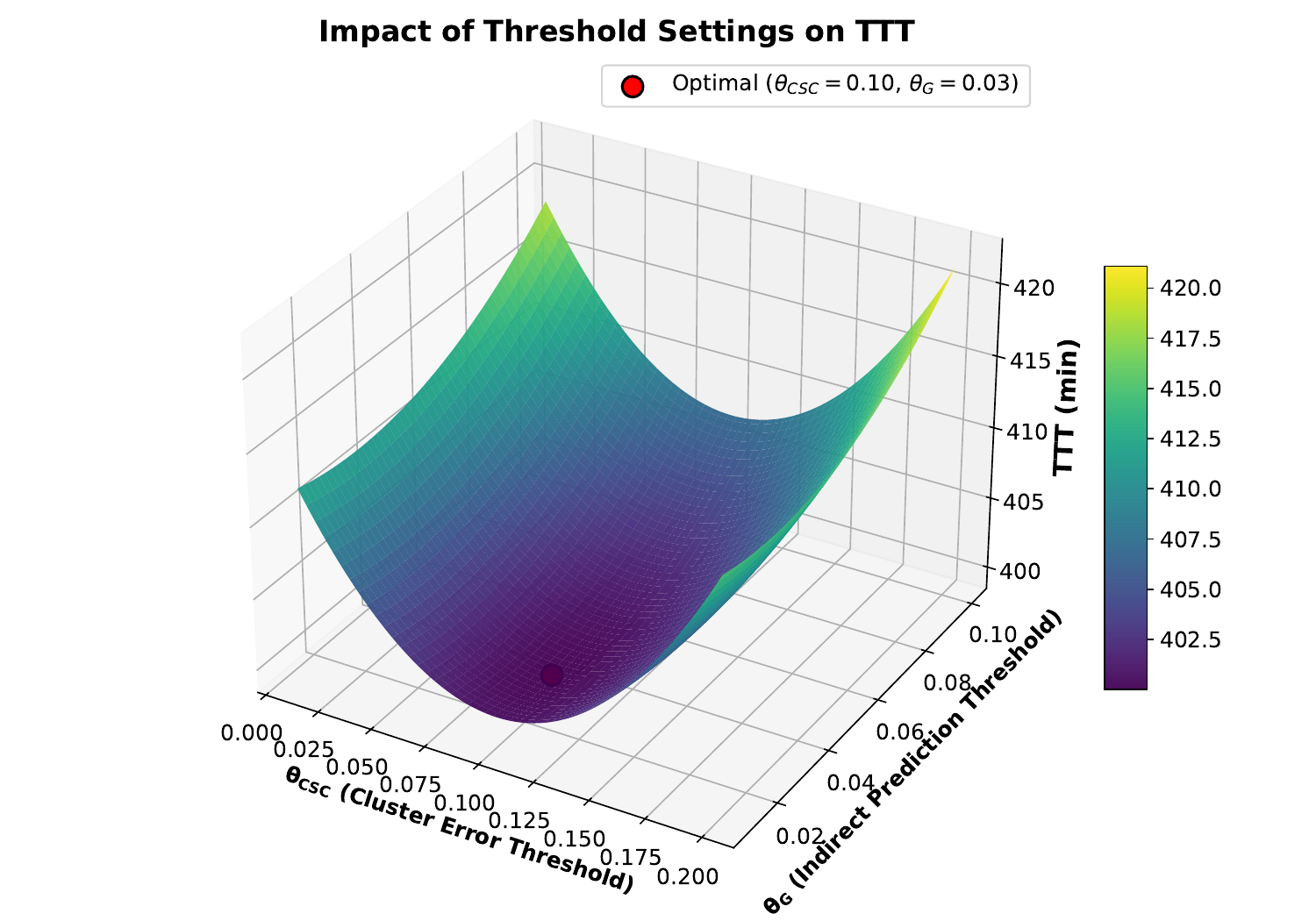}
}
\caption{The impact of error threshold settings on runtime update process.}
\label{fig:param}
% \Description[Figure 5]{Impact of update. The x-axis 
% is the pre-defined target positions shown in \autoref{tb:TTT} and CVEs shown in \autoref{tb:TTE} respectively. The y-axis is the relative time consumption compared with \sys. }
\end{figure}

We evaluate $\theta_{CSC}$ over the range [0\%, 20\%] in 2.5\% increments, and $\theta_G$ over [0\%, 10\%] in 0.5\% increments. Each configuration is tested using the UniBench benchmark suite, where we measure the average Time-to-Target (TTT) across all target programs. As shown in \autoref{fig:param}, the configuration $\theta_{CSC}=10\%$ and $\theta_G=3\%$ consistently achieves the best trade-off — minimizing unnecessary updates while remaining responsive to prediction drift. This setting yields the lowest average TTT, balancing guidance quality with computational efficiency.

Thresholds below this range lead to overly frequent updates, which increase computational cost and reduce effective fuzzing throughput. Conversely, higher thresholds delay necessary updates, causing stale feasibility scores that misguide the fuzzer. While the defaults provide a robust general-purpose configuration, \sys also exposes both parameters for domain-specific tuning, allowing users to optimize the trade-off between accuracy and resource efficiency based on workload characteristics.

\section{Limitations}

While \sys demonstrates substantial improvements over prior directed fuzzing techniques, it also has several limitations that merit further exploration. We outline three primary limitations observed during our evaluation.

\PP{Runtime Overhead for Feasibility Assignment.}  
AFLGopher introduces a modest runtime overhead during the initial phase of fuzzing, stemming from the need to collect execution traces and compute feasibility-aware distance metrics. This initialization phase typically lasts approximately five minutes before feasibility-informed guidance becomes effective. While this overhead is negligible for long-running fuzzing campaigns, it may hinder responsiveness in time-sensitive settings or programs with short execution windows. For instance, in the UniBench targets \textsc{cflow:parser.c:1223} and \textsc{cflow:parser.c:108}, as well as CVE-2016-4487 from the ATS suite, we observed a measurable delay in reaching the target due to this "warm-up" latency. Importantly, this initialization cost is incurred post-fuzzing launch and is fully accounted for in the evaluation metrics, ensuring that all reported speedups reflect end-to-end performance. Despite this delay, AFLGopher consistently outperforms state-of-the-art baselines across the majority of benchmarks. Nonetheless, optimizing early trace collection and reducing cold-start latency remain promising directions for future work, particularly to enhance applicability in real-time fuzzing pipelines or continuous integration environments.

% \PP{Prioritization of Feasible Paths over Exploitable Ones.}  
% \sys is designed to guide fuzzing along the most feasible paths toward target locations. However, the most feasible path is not always the most vulnerable one. In certain cases—such as CVE-2017-9047—\sys prioritizes high-feasibility paths that delay exposure of the actual vulnerability. While this is a deliberate trade-off to reduce wasted exploration, future work could explore integrating exploitability metrics (e.g., crash dump features~\cite{b42}) into the feasibility model to better balance coverage and vulnerability exposure.

\PP{Resource Cost of Runtime Updates.}  
To maintain adaptive guidance, \sys periodically evaluates prediction accuracy and updates feasibility models. Although these updates are conducted asynchronously, they still consume memory and CPU resources. In environments with constrained resources or high mutation throughput, this can lead to slight performance degradation. For example, in case \textsc{mp3gain:layer3.c:1116} from \autoref{tb:TTT}, \sys underperforms compared to SelectFuzz, which aggressively explores all reachable paths without feasibility modeling. While \sys’s several updates in this case enhanced model precision, they also incurred non-trivial overhead. Currently, update frequency is controlled via error thresholds (see \autoref{sec:impact of update}); more intelligent scheduling strategies or lightweight approximations may offer improvements in future versions.

\PP{Summary}  
Despite these limitations, \sys consistently delivers superior performance across a wide range of targets. Moreover, each limitation presents a clear path toward further enhancement—through trace optimization, feasibility-exploitability fusion, or smarter update scheduling—reinforcing the system’s extensibility and relevance to future work in directed fuzzing.

\section{Related Work}

\PP{Directed Grey-Box Fuzzing}  
Directed grey-box fuzzing (DGF) has gained increasing traction for its ability to efficiently test specific program regions, particularly those associated with vulnerabilities or patches. AFLGo~\cite{b3} introduced the foundational concept of DGF by using control-flow distances to prioritize seeds closer to target locations. Hawkeye~\cite{b2} builds upon this approach by improving seed selection and incorporating indirect-call targets, resulting in more complete path coverage. 

Several other systems have proposed alternative methods to direct fuzzing toward program hotspots. Liang et al.~\cite{b4} presented a lightweight sequence-coverage–based technique for user-defined statement targeting. Wüstholz et al.~\cite{b10} employed online static analysis to refine directionality. UAFuzz~\cite{b12} focuses on use-after-free vulnerabilities at the binary level, becoming the first directed fuzzer specialized for such a class.

Recent state-of-the-art tools include WindRanger~\cite{b54}, which uses deviation blocks to guide distance computation; Beacon~\cite{b61}, which performs lightweight static analysis for path pruning; and SelectFuzz~\cite{b42}, which selectively instruments paths most relevant to the target site. Other noteworthy systems like DeepGo~\cite{b70}, HyperGo~\cite{b67}, and DAFL~\cite{b68} also contribute meaningful advances, particularly in enhancing guidance through learned metrics. However, DeepGo and HyperGo are unavailable for reproduction, and DAFL's integration presented practical challenges. As a result, we evaluate against WindRanger, Beacon, and SelectFuzz as representative, open-source baselines.

To the best of our knowledge, none of these systems incorporates feasibility-aware guidance. In contrast, \sys is the first to leverage both branch and indirect-call feasibility—learned dynamically and semantically grouped—to improve the precision of directed fuzzing.

\PP{Machine Learning–Assisted Fuzzing.}  
Recent years have also seen a rise in machine learning (ML) techniques applied to various aspects of fuzzing. Fuzzguard~\cite{b30} uses deep learning to predict input reachability, filtering out low-value seeds before execution. DeFuzz~\cite{b16} identifies vulnerable code regions via pre-trained models and directs fuzzing accordingly. DeepFuzz~\cite{b51} employs a sequence-to-sequence generator to create well-formed C programs for compiler fuzzing.

Neuzz~\cite{b43} applies surrogate neural models to smooth program branches and facilitate gradient-guided input generation. Angora~\cite{b49} similarly relies on gradient descent to solve path constraints, improving branch coverage. Suzzer~\cite{b38} prioritizes paths statistically more likely to contain bugs. Meanwhile, RL-based approaches such as Wang et al.~\cite{b45} introduce hierarchical schedulers for coverage-maximizing seed selection.

These systems predominantly focus on improving input mutation strategies or identifying coverage-critical paths. In contrast, \sys contributes a novel dimension to ML-assisted fuzzing by:  
(1) clustering conditional statements via unsupervised learning to support generalizable feasibility estimation, and  
(2) using an LSTM-based model to predict the runtime reachability of indirect-call targets. Rather than enhancing input generation, \sys enhances guidance precision, thereby improving the fuzzing process at the semantic and structural levels.

\section{Conclusion}

Directed grey-box fuzzing often struggles with inefficiencies due to feasibility-unaware distance calculations. In this paper, we introduce a feasibility-aware approach with \sys, a tool that predicts the feasibility of branch statements and incorporates this feedback into the fuzzing. \sys combines branch-statement classification with an efficient model update mechanism to ensure precision and efficiency. In performance evaluations using benchmark datasets, \sys significantly outperformed state-of-the-art directed fuzzers in both Time-To-Target and Time-To-Exposure metrics. These results confirm the superior effectiveness of \sys over contemporary directed fuzzers.

\bibliographystyle{ACM-Reference-Format}
\bibliography{main}

\appendix
\label{sec:appendix}

\begin{table*}
    \centering
        \centering
        \caption{The Time-to-Target Results on Programs from UniBench}
        \resizebox{\linewidth}{!}{
            \begin{tabular}{ccccccccccccc}
\toprule
\textbf{ID} &                \textbf{Program} &            \textbf{Version} &             \textbf{Target Position} &          \textbf{AFLGo} &         \textbf{BEACON} &     \textbf{WindRanger} &     \textbf{SelectFuzz} &        \textbf{AFLGo-ICall} &    \textbf{AFLGopher-C} &    \textbf{AFLGopher-N} &    \textbf{AFLGopher-U} &      \textbf{AFLGopher} \\ \toprule

  1 &              \multirow{4}*{cflow} &               \multirow{4}*{1.6} &            parser.c:281 & \textit{T.O.} &         3.08h &         2.56h &         2.33h & \textit{T.O.} & \textit{T.O.} &        17.17h &         6.22h &         2.44h \\
  2 &                    &                   &                c.c:1783 &         1.10h &         0.15h &         0.11h &         0.11h &         0.30h &         0.48h &         0.64h &         0.55h &         0.20h \\
  3 &                    &                   &           parser.c:1223 &         0.12h &         0.07h &         0.11h &         2.76h &         0.13h &         0.40h &         0.44h &         0.43h &         0.16h \\
  4 &                    &                   &            parser.c:108 &         1.30h &         2.70h &         0.43h &         0.45h &         1.20h &         0.92h &         1.27h &         0.87h &         0.34h \\ \midrule
  5 &            \multirow{4}*{mp42aac} &  \multirow{4}*{Bento4 1.5.1-628} &      Ap4AvccAtom.cpp:82 & \textit{T.O.} &           N/A & \textit{T.O.} & \textit{T.O.} & \textit{T.O.} &         8.47h &        10.10h &         8.15h &         3.49h \\
  6 &                    &                   &     Ap4TrunAtom.cpp:139 & \textit{T.O.} &           N/A & \textit{T.O.} &         9.22h & \textit{T.O.} &        17.28h &        20.72h &        18.71h &         6.95h \\
  7 &                    &                   &     Ap4SbgpAtom.cpp:181 & \textit{T.O.} &           N/A & \textit{T.O.} &         0.80h & \textit{T.O.} &        21.21h & \textit{T.O.} &        21.49h &         8.74h \\
  8 &                    &                   &  Ap4AtomFactory.cpp:490 & \textit{T.O.} &           N/A & \textit{T.O.} &        12.57h & \textit{T.O.} &        11.05h &        12.72h &         9.53h &         4.31h \\
  \midrule
  9 &              \multirow{4}*{jhead} &                 \multirow{4}*{3.0.0} &             exif.c:1339 & \textit{T.O.} &           N/A & \textit{T.O.} & \textit{T.O.} & \textit{T.O.} &        22.08h &        22.93h &        18.43h &         7.95h \\
 10 &                    &                   &              iptc.c:143 & \textit{T.O.} &           N/A & \textit{T.O.} & \textit{T.O.} & \textit{T.O.} &        21.44h &        23.65h &        22.67h &         7.88h \\
 11 &                    &                   &               iptc.c:91 & \textit{T.O.} &           N/A & \textit{T.O.} & \textit{T.O.} & \textit{T.O.} &         5.31h & \textit{T.O.} &         5.26h &         2.18h \\
 12 &                    &                   &         makernote.c:174 & \textit{T.O.} &           N/A & \textit{T.O.} & \textit{T.O.} & \textit{T.O.} &        16.95h &        19.97h &        18.98h &         6.78h \\
 \midrule
 13 &            \multirow{4}*{mp3gain} &             \multirow{4}*{1.5.2} &           layer3.c:1116 &        23.80h &           N/A & \textit{T.O.} &         6.76h & \textit{T.O.} &        20.73h &        23.23h &        17.13h &         7.74h \\
 14 &                    &                   &           mp3gain.c:602 & \textit{T.O.} &           N/A & \textit{T.O.} & \textit{T.O.} & \textit{T.O.} &         5.52h &         7.19h &         5.03h &         2.19h \\
 15 &                    &                   &         interface.c:663 & \textit{T.O.} &           N/A & \textit{T.O.} &         2.52h & \textit{T.O.} &         7.38h & \textit{T.O.} &         6.93h &         2.72h \\
 16 &                    &                   &            apetag.c:341 &        21.55h &           N/A &         8.10h & \textit{T.O.} &        17.63h & \textit{T.O.} & \textit{T.O.} & \textit{T.O.} &        14.66h \\
 \midrule
 17 &               \multirow{4}*{lame} &            \multirow{4}*{3.99.5} &         bitstream.c:823 & \textit{T.O.} &           N/A & \textit{T.O.} & \textit{T.O.} & \textit{T.O.} &        15.25h &        18.61h &        14.55h &         6.09h \\
 18 &                    &                   &             lame.c:2148 & \textit{T.O.} &           N/A & \textit{T.O.} &        11.89h &        18.73h & \textit{T.O.} & \textit{T.O.} & \textit{T.O.} &        11.13h \\
 19 &                    &                   &       uantize\_pvt.c:441 & \textit{T.O.} &           N/A &        14.15h & \textit{T.O.} & \textit{T.O.} &        15.47h &        17.81h &        16.74h &         6.20h \\
 20 &                    &                   &        get\_audio.c:1605 & \textit{T.O.} &           N/A & \textit{T.O.} &         9.09h & \textit{T.O.} &        22.96h & \textit{T.O.} & \textit{T.O.} &         9.21h \\
 \midrule
 21 &            \multirow{4}*{imginfo} &     \multirow{4}*{jasper 2.0.1} &           jp2\_cod.c:841 & \textit{T.O.} &           N/A & \textit{T.O.} & \textit{T.O.} & \textit{T.O.} & \textit{T.O.} & \textit{T.O.} &        22.28h &        10.10h \\
 22 &                    &                   &           jp2\_cod.c:636 & \textit{T.O.} &           N/A & \textit{T.O.} &         2.97h &        14.71h &        18.08h &        19.64h &        15.29h &         6.87h \\
 23 &                    &                   &        jas\_stream.c:823 & \textit{T.O.} &           N/A &        16.40h & \textit{T.O.} & \textit{T.O.} &        12.34h &        14.91h &        12.38h &         4.86h \\
 24 &                    &                   &          jpc\_dec.c:1393 & \textit{T.O.} &           N/A & \textit{T.O.} &        14.50h & \textit{T.O.} &        12.28h &        15.95h &        12.63h &         5.06h \\
 \midrule
 25 & \multirow{4}*{gdk-pixbuf-pixdata} & \multirow{4}*{gdk-pixbuf 2.31.1} & gdk-pixbuf-loader.c:387 & \textit{T.O.} &           N/A &           N/A &           N/A & \textit{T.O.} &        10.18h &        11.11h &         9.07h &         3.96h \\
 26 &                    &                   &           io-qtif.c:511 & \textit{T.O.} &           N/A &           N/A &           N/A &        14.72h &         5.10h &         6.73h &         4.97h &         2.10h \\
 27 &                    &                   &           io-jpeg.c:691 & \textit{T.O.} &           N/A &           N/A &           N/A & \textit{T.O.} &         7.33h &         9.40h &         8.60h &         2.95h \\
 28 &                    &                   &            io-tga.c:360 &        21.29h &           N/A &           N/A &           N/A &        20.06h &        19.89h &        18.42h &         0.96h &         0.40h \\
\midrule
 29 &                 \multirow{4}*{jq} &               \multirow{4}*{1.5} &          jv\_dtoa.c:3122 & \textit{T.O.} &           N/A &           N/A &           N/A &         4.44h &        17.85h &        23.65h &        18.19h &         7.36h \\

 30 &                    &                   &          jv\_dtoa.c:2004 & \textit{T.O.} &           N/A &           N/A &           N/A & \textit{T.O.} &        23.53h & \textit{T.O.} &        22.45h &         8.75h \\
 31 &                    &                   &          jv\_dtoa.c:2518 & \textit{T.O.} &           N/A &           N/A &           N/A & \textit{T.O.} &        11.86h &        12.92h &        11.64h &         4.52h \\
 32 &                    &                   &         jv\_unicode.c:42 & \textit{T.O.} &           N/A &           N/A &           N/A &         4.60h &        16.89h &        19.61h &        16.55h &         6.73h \\
  \midrule
 33 &            \multirow{4}*{tcpdump} &             \multirow{4}*{4.8.1} &        print-aodv.c:259 & \textit{T.O.} &           N/A &        16.05h &           N/A & \textit{T.O.} &        12.69h &        15.82h &        11.48h &         4.88h \\
 34 &                    &                   &         print-ntp.c:412 & \textit{T.O.} &           N/A &        12.23h &           N/A & \textit{T.O.} &         7.83h &         8.48h &         7.85h &         2.89h \\

 35 &                    &                   &       print-rsvp.c:1252 & \textit{T.O.} &           N/A &        14.07h &           N/A & \textit{T.O.} &         5.90h & \textit{T.O.} &         5.15h &         2.32h \\
 36 &                    &                   &        print-l2tp.c:606 & \textit{T.O.} &           N/A & \textit{T.O.} &           N/A &         4.68h &        11.50h &        13.63h &        12.44h &         4.43h \\
  \midrule
 37 &                \multirow{4}*{tic} &       \multirow{4}*{ncurses 6.1} &         captoinfo.c:189 & \textit{T.O.} &           N/A &           N/A &           N/A &         4.65h &         7.13h &         7.72h &         5.89h &         2.67h \\
 38 &                    &                   &       alloc entry.c:141 & \textit{T.O.} &           N/A &           N/A &           N/A & \textit{T.O.} &        14.42h &        17.22h &        12.51h &         5.41h \\
 39 &                    &                   &        name match.c:111 & \textit{T.O.} &           N/A &           N/A &           N/A & \textit{T.O.} &         3.78h &         4.61h &         3.72h &         1.45h \\

 40 &                    &                   &            entries.c:78 & \textit{T.O.} &           N/A &           N/A &           N/A & \textit{T.O.} &         9.81h &        11.28h &         8.90h &         3.75h \\
  \midrule
 41 &            \multirow{4}*{flvmeta} &             \multirow{4}*{1.2.1} &             json.c:1036 & \textit{T.O.} &           N/A & \textit{T.O.} & \textit{T.O.} &         4.45h &         9.39h &        12.07h &        11.34h &         3.85h \\
 42 &                    &                   &               api.c:718 & \textit{T.O.} &           N/A & \textit{T.O.} & \textit{T.O.} &         4.41h &        16.66h &        21.72h &        18.26h &         6.69h \\
 43 &                    &                   &          flvmeta.c:1023 & \textit{T.O.} &           N/A & \textit{T.O.} &         5.67h & \textit{T.O.} & \textit{T.O.} & \textit{T.O.} &        23.29h &         8.36h \\
 44 &                    &                   &             check.c:769 & \textit{T.O.} &           N/A & \textit{T.O.} & \textit{T.O.} &         4.76h &         9.39h &        12.44h &         8.99h &         3.88h \\
 \midrule
 45 &          \multirow{4}*{tiffsplit} &     \multirow{4}*{libtiff 3.9.7} &        tif\_ojpeg.c:1277 & \textit{T.O.} &           N/A & \textit{T.O.} &         2.54h & \textit{T.O.} &         7.33h &         8.76h &         6.74h &         2.95h \\
 46 &                    &                   &          tif\_read.c:335 & \textit{T.O.} &           N/A & \textit{T.O.} &         1.58h & \textit{T.O.} &        11.80h &        13.56h &         9.65h &         4.38h \\
 47 &                    &                   &          tif\_jbig.c:277 & \textit{T.O.} &           N/A & \textit{T.O.} &        11.70h & \textit{T.O.} & \textit{T.O.} & \textit{T.O.} & \textit{T.O.} &        11.87h \\
 48 &                    &                   &      tif\_dirread.c:1977 & \textit{T.O.} &           N/A & \textit{T.O.} & \textit{T.O.} & \textit{T.O.} & \textit{T.O.} & \textit{T.O.} & \textit{T.O.} &        13.62h \\
  \midrule
 49 &                 \multirow{4}*{nm} &  \multirow{4}*{binutils-5279478} &            tekhex.c:325 & \textit{T.O.} &        23.46h & \textit{T.O.} & \textit{T.O.} & \textit{T.O.} &         7.72h &         8.51h &         6.87h &         2.98h \\

 50 &                    &                   &              elf.c:8793 & \textit{T.O.} &        23.67h & \textit{T.O.} &        10.75h &         4.63h &        18.56h &        23.27h &        19.10h &         7.29h \\
 51 &                    &                   &           dwarf2.c:2378 & \textit{T.O.} &        18.51h & \textit{T.O.} &        11.68h & \textit{T.O.} &        10.35h &        12.14h &         9.35h &         3.76h \\
 52 &                    &                   &     elf-properties.c:51 & \textit{T.O.} & \textit{T.O.} & \textit{T.O.} & \textit{T.O.} & \textit{T.O.} & \textit{T.O.} & \textit{T.O.} & \textit{T.O.} &        10.19h \\
  \midrule
 53 &          \multirow{4}*{pdftotext} &                 \multirow{4}*{4.00} &             XRef.cc:645 & \textit{T.O.} &           N/A & \textit{T.O.} & \textit{T.O.} & \textit{T.O.} & \textit{T.O.} & \textit{T.O.} & \textit{T.O.} &        12.70h \\
 54 &                    &                   &         GfxFont.cc:1337 & \textit{T.O.} &           N/A & \textit{T.O.} & \textit{T.O.} &         4.76h &        12.95h &        17.19h &        11.84h &         5.34h \\

 55 &                    &                   &          Stream.cc:1004 & \textit{T.O.} &           N/A & \textit{T.O.} &         9.54h &         4.46h & \textit{T.O.} & \textit{T.O.} &        54.03h & \textit{T.O.} \\
 56 &                    &                   &         GfxFont.cc:1643 & \textit{T.O.} &           N/A & 
 \textit{T.O.} & \textit{T.O.} & \textit{T.O.} & \textit{T.O.} &        22.30h & \textit{T.O.} &        12.98h \\
  \midrule
 57 &            \multirow{4}*{sqlite3} &      \multirow{4}*{SQLite 3.8.9} &            pager.c:5017 & \textit{T.O.} &           N/A &           N/A & \textit{T.O.} & \textit{T.O.} &         8.07h &         9.55h &         6.93h &         3.14h \\
 58 &                    &                   &             func.c:1029 & \textit{T.O.} &           N/A &           N/A & \textit{T.O.} &         4.48h &        11.03h &        12.41h &        10.35h &         4.26h \\
 59 &                    &                   &           insert.c:1498 & \textit{T.O.} &           N/A &           N/A & \textit{T.O.} &         4.46h &        20.22h &        22.47h &        22.71h &         7.98h \\

 60 &                    &                   &             vdbe.c:1984 & \textit{T.O.} &        16.40h &           N/A &         9.74h & \textit{T.O.} &        12.88h &        14.06h &        12.68h &         4.65h \\
  \midrule
 61 &              \multirow{4}*{exiv2} &              \multirow{4}*{0.26} &    tiffcomposite.cpp:82 & \textit{T.O.} &           N/A &         7.12h &         7.35h &         3.20h &         7.62h &         9.30h &         6.75h &         2.88h \\
 62 &                    &                   &   XMPMeta-Parse.cpp:847 &        23.08h &           N/A &         2.45h & \textit{T.O.} &         4.56h &        18.00h &        21.47h &         0.89h &         0.32h \\
 63 &                    &                   &    tiffvisitor.cpp:1044 & \textit{T.O.} &           N/A & \textit{T.O.} &        11.53h &         3.37h & \textit{T.O.} &         4.93h &         4.37h &         1.60h \\
 64 &                    &                   &   XMPMeta-Parse.cpp:896 & \textit{T.O.} &           N/A &        21.47h &        11.68h &         3.56h &         2.35h &         2.66h &         2.40h &         0.94h \\

  \midrule
 65 &            \multirow{4}*{objdump} &     \multirow{4}*{binutils-2.28} &              elf.c:9509 & \textit{T.O.} &        20.92h & \textit{T.O.} &         8.08h & \textit{T.O.} &        16.28h &        20.30h &        14.72h &         6.52h \\
 66 &                    &                   &           section.c:936 & \textit{T.O.} & \textit{T.O.} & \textit{T.O.} &         0.87h & \textit{T.O.} &        12.80h &        14.61h &        14.15h &         4.72h \\
 67 &                    &                   &              bfd.c:1108 & \textit{T.O.} &        23.48h & \textit{T.O.} & \textit{T.O.} & \textit{T.O.} &         7.59h &         8.42h &         6.86h &         2.81h \\
 68 &                    &                   &             bfdio.c:262 & \textit{T.O.} & \textit{T.O.} & \textit{T.O.} & \textit{T.O.} & \textit{T.O.} &        12.10h &        15.88h &        12.46h &         4.95h \\
  \midrule
 69 &             \multirow{4}*{ffmpeg} &             \multirow{4}*{4.0.1} &            rawdec.c:268 & \textit{T.O.} &           N/A & \textit{T.O.} &         5.06h &         4.51h &        14.49h &        17.18h &        14.06h &         5.99h \\

 70 &                    &                   &            decode.c:557 & \textit{T.O.} &           N/A & \textit{T.O.} &        13.66h &         4.58h & \textit{T.O.} &        17.91h &         3.52h &         1.29h \\
 71 &                    &                   &              dump.c:632 & \textit{T.O.} &           N/A & \textit{T.O.} & \textit{T.O.} & \textit{T.O.} &         9.32h &        11.30h &        10.51h &         3.72h \\
 72 &                    &                   &                 utils.c & \textit{T.O.} &           N/A & \textit{T.O.} &        11.46h &         4.46h & \textit{T.O.} & \textit{T.O.} & \textit{T.O.} &         9.30h \\
 \midrule
 73 &               \multirow{4}*{mujs} &             \multirow{4}*{1.0.2} &             jsrun.c:572 & \textit{T.O.} &           N/A & \textit{T.O.} & \textit{T.O.} & \textit{T.O.} & \textit{T.O.} &        19.85h &        18.39h &         6.58h \\
 74 &                    &                   &               jsgc.c:47 & \textit{T.O.} &           N/A & \textit{T.O.} & \textit{T.O.} &         4.57h & \textit{T.O.} & \textit{T.O.} & \textit{T.O.} & \textit{T.O.} \\

 75 &                    &                   &            jsdump.c:292 & \textit{T.O.} &           N/A &        10.67h & \textit{T.O.} & \textit{T.O.} & \textit{T.O.} & \textit{T.O.} &        21.60h &         9.78h \\
 76 &                    &                   &           jsvalue.c:362 & \textit{T.O.} &           N/A & \textit{T.O.} & \textit{T.O.} & \textit{T.O.} &        22.22h &         3.62h &         2.57h &         1.16h \\
  \midrule
 77 &           \multirow{4}*{swftools} &             \multirow{4}*{0.9.2} &          initcode.c:242 & \textit{T.O.} &           N/A &        18.53h &        12.22h & \textit{T.O.} &        23.46h & \textit{T.O.} & \textit{T.O.} &        11.55h \\
 78 &                    &                   &               png.c:410 & \textit{T.O.} &           N/A &        11.31h &        13.14h & \textit{T.O.} &        14.75h &         8.29h &         6.47h &         2.60h \\
 79 &                    &                   &              poly.c:137 & \textit{T.O.} &           N/A & \textit{T.O.} &         3.21h & \textit{T.O.} & \textit{T.O.} & \textit{T.O.} & \textit{T.O.} & \textit{T.O.} \\

 80 &                    &                   &          jpeg2swf.c:257 & \textit{T.O.} &           N/A &         9.19h &         8.32h &         4.65h &        20.43h & \textit{T.O.} &        19.27h &         7.66h \\
  \toprule
\multicolumn{4}{c}{\textbf{\textit{speedup}}} & \multicolumn{1}{c}{\textbf{3.76x}} & \multicolumn{1}{c}{\textbf{2.57x}} & \multicolumn{1}{c}{\textbf{3.30x}} & \multicolumn{1}{c}{\textbf{2.52x}} & \multicolumn{1}{c}{\textbf{2.86x}} & \multicolumn{1}{c}{\textbf{2.47x}} & \multicolumn{1}{c}{\textbf{2.69x}} & \multicolumn{1}{c}{\textbf{2.23x}} & \multicolumn{1}{c}{\textbf{-}} ~ \\

\multicolumn{4}{c}{\textbf{mean \textit{p-values}}} & \multicolumn{1}{c}{\textbf{0.009}} & \multicolumn{1}{c}{\textbf{0.004}} & \multicolumn{1}{c}{\textbf{0.017}} & \multicolumn{1}{c}{\textbf{0.005}} & \multicolumn{1}{c}{\textbf{0.012}} & \multicolumn{1}{c}{\textbf{0.03}} & \multicolumn{1}{c}{\textbf{0.008}} & \multicolumn{1}{c}{\textbf{0.014}} & \multicolumn{1}{c}{\textbf{-}} ~ \\

\multicolumn{4}{c}{\textbf{mean $\hat A_{12}$}} & \multicolumn{1}{c}{\textbf{0.78}} & \multicolumn{1}{c}{\textbf{0.81}} & \multicolumn{1}{c}{\textbf{0.86}} & \multicolumn{1}{c}{\textbf{0.82}} & \multicolumn{1}{c}{\textbf{0.79}} & \multicolumn{1}{c}{\textbf{0.79}} & \multicolumn{1}{c}{\textbf{0.83}} & \multicolumn{1}{c}{\textbf{0.80}} & \multicolumn{1}{c}{\textbf{-}} ~ \\
\bottomrule
\end{tabular}
 % Your first table data
        }
        \label{tb:TTT}
\end{table*}

\end{document}